
%
\magnification=1200
%
%
\hsize=31pc
\vsize=55 truepc
\hfuzz=2pt
\vfuzz=4pt
\pretolerance=5000
\tolerance=5000
\parskip=0pt plus 1pt
\parindent=16pt
%
%
\font\fourteenrm=cmr10 scaled \magstep2
\font\fourteeni=cmmi10 scaled \magstep2
\font\fourteenbf=cmbx10 scaled \magstep2
\font\fourteenit=cmti10 scaled \magstep2
\font\fourteensy=cmsy10 scaled \magstep2
\font\large=cmbx10 scaled \magstep1
%

%
%

%
%
\font\bdi=cmmib10
%
%

%
%
\font\eightrm=cmr8
\font\eighti=cmmi8
\font\eightbf=cmbx8
\font\eightit=cmti8

\font\eightsy=cmsy8
\font\sixrm=cmr6
\font\sixi=cmmi6
\font\sixsy=cmsy6

\def\tenpoint{\def\rm{\fam0\tenrm}%
  \textfont0=\tenrm \scriptfont0=\sevenrm
                      \scriptscriptfont0=\fiverm
  \textfont1=\teni  \scriptfont1=\seveni
                      \scriptscriptfont1=\fivei
  \textfont2=\tensy \scriptfont2=\sevensy
                      \scriptscriptfont2=\fivesy
  \textfont3=\tenex   \scriptfont3=\tenex
                      \scriptscriptfont3=\tenex
  \textfont\itfam=\tenit  \def\it{\fam\itfam\tenit}%
  \textfont\slfam=\tensl  \def\sl{\fam\slfam\tensl}%
  \textfont\bffam=\tenbf  \scriptfont\bffam=\sevenbf
                            \scriptscriptfont\bffam=\fivebf
                            \def\bf{\fam\bffam\tenbf}%
  \normalbaselineskip=20 truept
  \setbox\strutbox=\hbox{\vrule height14pt depth6pt
width0pt}%
  \let\sc=\eightrm \normalbaselines\rm}
\def\eightpoint{\def\rm{\fam0\eightrm}%
  \textfont0=\eightrm \scriptfont0=\sixrm
                      \scriptscriptfont0=\fiverm
  \textfont1=\eighti  \scriptfont1=\sixi
                      \scriptscriptfont1=\fivei
  \textfont2=\eightsy \scriptfont2=\sixsy
                      \scriptscriptfont2=\fivesy
  \textfont3=\tenex   \scriptfont3=\tenex
                      \scriptscriptfont3=\tenex
  \textfont\itfam=\eightit  \def\it{\fam\itfam\eightit}%
  \textfont\bffam=\eightbf  \def\bf{\fam\bffam\eightbf}%
  \normalbaselineskip=16 truept
  \setbox\strutbox=\hbox{\vrule height11pt depth5pt width0pt}}
\def\fourteenpoint{\def\rm{\fam0\fourteenrm}%
  \textfont0=\fourteenrm \scriptfont0=\tenrm
                      \scriptscriptfont0=\eightrm
  \textfont1=\fourteeni  \scriptfont1=\teni
                      \scriptscriptfont1=\eighti
  \textfont2=\fourteensy \scriptfont2=\tensy
                      \scriptscriptfont2=\eightsy
  \textfont3=\tenex   \scriptfont3=\tenex
                      \scriptscriptfont3=\tenex
  \textfont\itfam=\fourteenit  \def\it{\fam\itfam\fourteenit}%
  \textfont\bffam=\fourteenbf  \scriptfont\bffam=\tenbf
                             \scriptscriptfont\bffam=\eightbf
                             \def\bf{\fam\bffam\fourteenbf}%
  \normalbaselineskip=24 truept
  \setbox\strutbox=\hbox{\vrule height17pt depth7pt width0pt}%
  \let\sc=\tenrm \normalbaselines\rm}

\def\today{\number\day\ \ifcase\month\or
  January\or February\or March\or April\or May\or June\or
  July\or August\or September\or October\or November\or
December\fi
  \space \number\year}
%
%
\newcount\secno      
\newcount\subno      
\newcount\subsubno   
\newcount\appno      
\newcount\tableno    
\newcount\figureno   
\normalbaselineskip=20 truept
\baselineskip=20 truept
%
%
\def\title#1
   {\vglue1truein
   {\baselineskip=24 truept
    \pretolerance=10000
    \raggedright
    \noindent \fourteenpoint\bf #1\par}
    \vskip1truein minus36pt}
%
%
\def\author#1
  {{\pretolerance=10000
    \raggedright
    \noindent {\large #1}\par}}
%
%
\def\address#1
   {\bigskip
    \noindent \rm #1\par}
%
%
\def\shorttitle#1
   {\vfill
    \noindent \rm Short title: {\sl #1}\par
    \medskip}
%
%
\def\pacs#1
   {\noindent \rm PACS number(s): #1\par
    \medskip}
%
%
\def\jnl#1
   {\noindent \rm Submitted to: {\sl #1}\par
    \medskip}
%
%
\def\date
   {\noindent Date: \today\par
    \medskip}
%
%
\def\beginabstract
   {\vfill\eject
    \noindent {\bf Abstract. }\rm}
%
%
\def\keyword#1
   {\bigskip
    \noindent {\bf Keyword abstract: }\rm#1}
%
%
\def\endabstract
   {\par
    \vfill\eject}
%
%
%

%
%
\def\entry#1#2#3
   {\noindent
    \hangindent=20pt
    \hangafter=1
    \hbox to20pt{#1 \hss}#2\hfill #3\par}
%
%
\def\subentry#1#2#3
   {\noindent
    \hangindent=40pt
    \hangafter=1
    \hskip20pt\hbox to20pt{#1 \hss}#2\hfill #3\par}
%
%
\def\section#1
   {\vskip0pt plus.1\vsize\penalty-250
    \vskip0pt plus-.1\vsize\vskip24pt plus12pt minus6pt
    \subno=0 \subsubno=0
    \global\advance\secno by 1
    \noindent {\bf \the\secno. #1\par}
    \bigskip
    \noindent}
%
%
\def\subsection#1
   {\vskip-\lastskip
    \vskip24pt plus12pt minus6pt
    \bigbreak
    \global\advance\subno by 1
    \subsubno=0
    \noindent {\sl \the\secno.\the\subno. #1\par}
    \nobreak
    \medskip
    \noindent}
%
%
\def\subsubsection#1
   {\vskip-\lastskip
    \vskip20pt plus6pt minus6pt
    \bigbreak
    \global\advance\subsubno by 1
    \noindent {\sl \the\secno.\the\subno.\the\subsubno. #1}\null. }
%
%
\def\appendix#1
   {\vskip0pt plus.1\vsize\penalty-250
    \vskip0pt plus-.1\vsize\vskip24pt plus12pt minus6pt
    \subno=0
    \global\advance\appno by 1
    \noindent {\bf Appendix \the\appno. #1\par}
    \bigskip
    \noindent}
%
%
\def\subappendix#1
   {\vskip-\lastskip
    \vskip36pt plus12pt minus12pt
    \bigbreak
    \global\advance\subno by 1
    \noindent {\sl \the\appno.\the\subno. #1\par}
    \nobreak
    \medskip
    \noindent}
%
%

%
%

%
%
\def\tabcaption#1
   {\global\advance\tableno by 1
    \noindent {\bf Table \the\tableno.} \rm#1\par
    \bigskip}
%
%
\def\figures
   {\vfill\eject
    \noindent {\bf Figure captions\par}
    \bigskip}
%
%
\def\figcaption#1
   {\global\advance\figureno by 1
    \noindent {\bf Figure \the\figureno.} \rm#1\par
    \bigskip}
%
%
\def\references
     {\vfill\eject
     {\noindent \bf References\par}
      \parindent=0pt
      \bigskip}
%
%
\def\refjl#1#2#3#4
   {\hangindent=16pt
    \hangafter=1
    \rm #1
   {\frenchspacing\sl #2
    \bf #3}
    #4\par}
%
%
\def\refbk#1#2#3
   {\hangindent=16pt
    \hangafter=1
    \rm #1
   {\frenchspacing\sl #2}
    #3\par}
%
%
\def\numrefjl#1#2#3#4#5
   {\parindent=40pt
    \hang
    \noindent
    \rm {\hbox to 30truept{\hss #1\quad}}#2
   {\frenchspacing\sl #3\/
    \bf #4}
    #5\par\parindent=16pt}
%
%
\def\numrefbk#1#2#3#4
   {\parindent=40pt
    \hang
    \noindent
    \rm {\hbox to 30truept{\hss #1\quad}}#2
   {\frenchspacing\sl #3\/}
    #4\par\parindent=16pt}
%
%

%
%
\def\frac#1#2{{#1 \over #2}}
%
%

%
%
\def\d{{\rm d}}
%
%
\def\e{{\rm e}}
%
%
\def\i{\ifmmode{\rm i}\else\char"10\fi}
%
%

%
%

%
%

%
%
\def\etal{{\sl et al\/}\ }
%
%
\catcode`\@=11
%
%
\def\ind{\hbox to 5pc{}}
%
%
\def\eq(#1){\hfill\llap{(#1)}}
%
%

%
%
\def\deqn#1{\displ@y\halign{\hbox to \displaywidth
    {$\@lign\displaystyle##\hfil$}\crcr #1\crcr}}
%
%
\def\indeqn#1{\displ@y\halign{\hbox to \displaywidth
    {$\ind\@lign\displaystyle##\hfil$}\crcr #1\crcr}}
%
%
\def\indalign#1{\displ@y \tabskip=0pt
  \halign to\displaywidth{\ind$\@lign\displaystyle{##}$\tabskip=0pt
    &$\@lign\displaystyle{{}##}$\hfill\tabskip=\centering
    &\llap{$\@lign##$}\tabskip=0pt\crcr
    #1\crcr}}
\catcode`\@=12
%
%



%
%


\def\NL{Nonlinearity}

%
%

\def\PR{Phys. Rev.}
\def\PRL{Phys. Rev. Lett.}

%
%

\def\bid#1{\hbox{$#1\/ \textfont1=\bdi$}}      
\font\extsyten=lasy10
\textfont10=\extsyten
\mathchardef\Box="2A32
\def\\{\hfill\break}
\def\dx{\delta x}
\def\xi{x_i}
\def\ga{\gamma}
\def\gu{\alpha}
\def\gd{\beta}
\def\la{\lambda}
\def\lc{l_{\rm c}}
\def\Lap#1{\widetilde{#1}}

\def\avt#1{\left\langle#1\right\rangle}
\def\ztq{\avt{|\widetilde{z}_i^{(j)}|^2}}

\title{One-dimensional asymmetrically coupled maps with defects}

\author{L Biferale\ddag, A Crisanti\dag, M Falcioni\dag\ and A Vulpiani\dag}

\address{\dag\ Dipartimento di Fisica, Universit\`a di Roma
                              ``La Sapienza'', I--00185 Roma, Italy}
\address{\ddag\ Dipartimento di Fisica, Universit\`a di Roma
                              ``Tor Vergata'', I--00133 Roma, Italy}

\shorttitle{chaos in extended systems}


\date

\beginabstract

In this letter we study chaotic dynamical properties of an asymmetrically
coupled one-dimensional chain of maps. We discuss the existence of coherent
regions in terms of the presence of defects along the chain. We find out
that temporal chaos is instantaneously localized around one single defect
and that the tangent vector jumps from one defect to another in an
apparently random way. We quantitatively measure the localization properties
by defining an entropy-like function in the space of tangent vectors.
\endabstract

Spatiotemporal chaos has been recently studied in many fields, as B\'enard
convection, optical turbulence, chemical reaction-diffusion systems, and so
on [1]. One of the paradigms of spatial chaos is the chaotic evolution of a
spatial pattern. Order and chaos are often considered to be opposite notions
in nature. One may think of order as a stable and regular behaviour, and of
chaos as unstable and erratic behaviour. Then one may be tempted to conclude
that spatial order cannot exist in temporally chaotic systems. There are,
however, many examples where spatial order coexists with temporal chaos,
e.g. convection, boundary layer, shear flows.

Generally, it is difficult to study this spatio-temporal behaviour from the
full equations, e.g. the Navier Stokes equations in fluids. As a consequence,
to gain more insight and to facilitate computations, model systems with a set
of coupled dynamical systems, coupled-map lattice, have been introduced. These
are a crude, but nontrivial, approximation of extended systems with discrete
space and time, but continuous states [1]. The simplest example is given by a
set of $N$ continuous variables $x_i$ which evolve in (discrete) time as
   $$\eqalign{
              x_i(n+1)&= (1-\gu_i-\gd_i)\, f(x_i(n)) \cr
           &\phantom{=}
             + \gu_i f(x_{i-1}(n)) + \gd_i f(x_{i+1}(n)). \cr}
    \eqno(1)$$
Usually, the coupling constants are assumed to be independent of the site
$i$ and equal, i.e. $\gu_i=\gd_i=\ga$, and periodic boundary conditions
are taken.

In some physical problems, e.g., shear flow, boundary layers or convection,
there is a privileged direction. This can be introduced in the model (1) by
taking asymmetric couplings [2,3]:
   $$ \gu_i=\gamma_1 \neq \gd_i=\gamma_2.
    \eqno(2) $$

The system (1)-(2) with periodic boundary conditions shows convective
instability [4]. In a reference frame moving with a constant velocity
$v$ in some band $[v_{\rm min},v_{\rm max}]$ an initial perturbation
$\dx_k(0)$ grows esponentially with a rate given by a co-moving Lyapunov
exponent $\lambda(v)$:

   $$\dx_{k+vn}(n)\simeq\dx_{k}(0)\, \e^{\lambda(v) n}.
    \eqno(3)$$

The situation becomes much more intriguing and interesting if one
introduces non-periodic boundary conditions. See, for example, Ref.~[2]
in which an unidirected lattice map is used to mimic convective
turbulence. Recently Aranson \etal [5] studied the system (1) with the
following coupling constants:
   $$\eqalign{
              \gu_i=& \gamma_1,\quad  \gd_i=\gamma_2 \quad
                       \hbox{\rm for}\ i=2,\dots,N-1 \cr
              \gu_1=& 0,\quad \gd_1= \gamma_2, \quad
              \gu_N=\gamma_1,\quad \gd_N=0, \cr
              \gamma_1 >& \gamma_2\cr}
    \eqno(4)$$
and $f(x)=R-x^2$ with $R=1.67$, for which the behaviour of a single map
is chaotic.

The system (1), (4) has the stable uniform solution,
   $$x_i(n)=\Lap{x}(n), \qquad
     \Lap{x}(n+1)=f(\Lap{x}(n)).
    \eqno(5)$$
Aranson \etal [5] found that starting from a randomly nonuniform initial
conditions, after some iterations the $x_i(n)$ become partially synchronized:
$x_i(n) \simeq \Lap{x}(n)$ for $i < \lc$, while for $i>\lc$ the $x_i$ are
spatially irregular. The finite coherence length $\lc$ is due to the
numerical noise. Indeed, they found that $\lc$ increases
logarithmically with the noise level in the numerical calculations.

This situation seems rather pathological, since the main features of a
chaotic system should not be destroyed by the presence of a small noise.

In order to gain intuition on this behaviour, we have studied the evolution
of the tangent vector of the system (1), (4). The evolution law of the
tangent vector $\bid{z}$ is obtained linearizing equation (1) about the
trajectory:
   $$\eqalign{
              z_i(n+1)&= (1-\gu_i-\gd_i)\, g(x_i(n))\, z_i(n) \cr
           &\phantom{=}
             + \gu_i g(x_{i-1}(n))\,z_{i-1}(n)
             + \gd_i g(x_{i+1}(n))\,z_{i+1}(n), \cr}
    \eqno(6)$$
where $g(y)=\d f(y)/ \d y$, and the coupling constants are given by (4).

A numerical analysis reveals that if the tangent vector is initially
localized in the  irregular zone of the system, i.e. $z_i(0)=0$ for
$i < \lc$, then one has a behaviour similar to the convective instability.
The vector translates, while growing, in the direction of increasing $i$.
However, unlike the case of periodic boundary conditions, as soon as the
perturbation reaches the boundary $i=N$, its amplitude decreases rapidly
to zero.

On the other hand, if $z_i(0)\neq 0$ for some $i < \lc$, the instability
does not travel and there is an exponential growth of $\bid{z}(n)$ in the
synchronized part of the chain. This means that the chaotic part of the
system is the spatially coherent one.

A measure of the degree of chaos is given by the Lyapunov exponents.
They measure the growth of the tangent vector for large time, and hence they
are ruled by the synchronized part of the system. As a consequence, since
$\lc$ depends on the noise level in the numerical calculations, we expect
that also the Lyapunov exponents should be affected. This is confirmed
by the numerical analysis. For example, in the case of $N=70$ we find only
one positive Lyapunov exponent, and the others negative, in a four-byte
precision calculation, while an eight-byte calculation leads to several
positive exponents.

As stated above, a similar scenario, is rather pathological in the framework
of chaotic systems, since one expects a sort of structural stability for the
main features. In our opinion the origin of this odd behaviour is in the open
boundary conditions. Indeed, for a one-dimensional system open boundary
condition represents a very strong ``defect''. The chain is broken somewhere.
For this reason we have studied a slightly modified version of the model which
in some sense interpolate between the periodic and the open boundary
conditions.

We consider the system (1)-(2), with $\gamma_1>\,\gamma_2$ and periodic
boundary conditions, and we introduce some ``defects'' by changing the coupling
constants at certain points $i=k_1,\cdots,k_M$ (with $M\ll N$) of the lattice.

We have considered the following class of defects:
   $$\gu_i=\gamma_1, \quad \gd_i=\gamma_2, \quad \hbox{\rm but}\quad
   \gu_i=\gamma_2, \quad \gd_i=\gamma_1, \quad \hbox{\rm if}\quad i=k_j
   .\eqno(7)$$

The dynamic evolution of a defects, i.e. of a variable with the exchanged
couplings, $x_{k_j}$, is essentially given by the single map solution
$\Lap{x}$, equation (5), as can be seen in a return map, see Fig.~1.
Thus the motion of the defects are not very much influenced by the other
variables. Yet there is a small synchronization length, $\lc$, of the
variables near the defects: $x_i(n)\simeq x_{k_j}(n)$ for $i-k_j \leq \lc$.
Unlike the case discussed above, this synchronization length  does not
depend on the noise in the numerical calculation. The doubling of the
precision in the computation does not produce any change on $\lc$.

In spite of this apparently simple behaviour, the dynamics is very
interesting if one studies the evolution in the tangent space. After a
short transient, independent on both $\bid{x}(0)$ and $\bid{z}(0)$, the
tangent vector $\bid{z}(n)$ get localized around one of the defects. It
stays there for a while, and then `jumps' on one of the two near neighbour
defects. It stays there for a while, and then jumps again, and so on. We
have, therefore, a sort of random walk of the position of the tangent
vector among the defects. In the meantime its modulus grows exponentially
revealing that the motion is chaotic.

One can generalize the analysis by considering the full set of Lyapunov
vectors $\bid{z}^{(j)}(n)$, with $j=1,\ldots,N$, associated with the
Lyapunov exponents $\la_j$. Each vector $\bid{z}^{(j)}(n)$ evolves
according to (6), but with orthogonal initial conditions, i.e.
$\bid{z}^{(j)}(0) \perp \bid{z}^{(j')}(0)$ if $j \neq j'$.
The Lyapunov exponents measure the exponential growth of the volume
individuated by the Lyapunov vectors. More explicitly,
   $$ \la _1 + \la _2 + \cdots + \la _q = \lim _{n\to \infty} \,
      {1 \over n} \ln {\vert \bid{z}^{(1)}(n)\land \bid{z}^{(2)}(n)
      \land \cdots \land \bid{z}^{(q)}(n)\vert \over
                       \vert \bid{z}^{(1)}(0)\land \bid{z}^{(2)}(0)
      \land \cdots \land \bid{z}^{(q)}(0)\vert}, \eqno(8)$$
where $\land$ indicates the external product.

The study of the time evolution of the tangent vectors for the system (1),
(7) reveals that $\bid{z}^{(j)}(n)$, for $j\leq M$ has the same
qualitative behaviour of $\bid{z}^{(1)}(n)$, while for $j>M$ does not
localize. Figure~2 shows $\ztq$ as a function of $i$, where
   $$ \ztq = \lim _{T\to \infty} {1\over T} \sum _{n=0}^{T-1} \,
      {\vert z_i^{(j)}(n)\vert ^2
      \over \sum _{l=1}^{N} \, \vert z_l^{(j)}(n)\vert ^2 }.
      \eqno(9)$$

A quantitative characterization of the degree of localization
may be obtained from the entropy [6]
   $$H^{(j)}= - \sum _{i=1}^{N} \,  \ztq \ln \ztq .\eqno(10)$$

We find that $H^{(j)}$, when $j\le M$, depends on the density of the
defects, $\rho = M/N$, and that there is a critical density
$\rho_{\rm c} \simeq \lc /N$. For $\rho < \rho_{\rm c}$ we have:
   $$ H^j-\ln N \sim c \ln \rho, \eqno(11)$$
where $c$ depends on $\gamma_1$ and $\gamma_2$, while for
$\rho > \rho_{\rm c}$ all the entropies $H^j$ saturate to a constant
value, see Fig.~3.

Finally we note that the Lyapunov exponents tend to cluster according to
the number of the defects, so that the spectrum $\la _i$ depends only on
the density of the defects, and exhibits the following behaviour
   $$ \la _i \simeq \la _1 h_{\rho} \left( i/N \right) ,
     \eqno(12)$$
where $\la _1$ depends only on the parameter of the local map and
$h_{\rho} (i/N)$ is a $\rho$-dependent function of $i/N$, see Fig.~4.
Formula (12) is well known [7] in the case of coupled maps without
asymmetric coupling and in the limit of large $N$.

Also these results do not depend on the noise level in the numerical
calculations.

{}From our analysis, by means of Lyapunov exponents and tangent vectors, we
have the following scenario. The system is driven by the defects, which
practically are not influenced by the rest of the system; moreover chaos
is concentrated on these defects, and the density of the defects is the
relevant property to determine the features of the Lyapunov exponents.
Something similar to a  deconfining transition concerning the localization
properties of chaos appears only for density of defects large enough to permit
considerable overlaps between the coherent regions of two adjacent defects.

We conclude by noting that the main features of this scenario hold also
for other classes of defects, e.g.:
   $$\gu_i=\gamma_1, \quad \gd_i=\gamma_2, \quad \hbox{\rm but}\quad
     \gu_i= \gd_i= 0 \quad \hbox{\rm if}\quad i=k_j .$$
This means that as long as  the chain is not broken (like in the
case of Aranson et al.) our results are quite robusts and everything is
independent of numerical precision.

\references
\numrefbk{[1]}{Crutchfield J P and Kaneko K 1987}{Directions in chaos}%
         {ed. Hao B L (Singapore: World Scientific) p 272}
\numrefbk{ }{Kaneko K 1990}{Formation, Dynamics and
              Statistics of Patterns}{ed.s Kawasaky K, Onnky A and
              Suzuki M (Singapore: World Scientific)}
\numrefjl{[2]}{Jensen M H 1989}{\PRL}{62}{1361}
\numrefjl{ }{Jensen M H 1989}{Physica D}{32}{203}
\numrefjl{[3]}{Aranson I, Afraimovich V S and Rabinovich M I 1990}%
         {\NL}{3}{639}
\numrefjl{[4]}{Bohr T and Rand A 1991}{Physica}{52D}{532}
\numrefjl{[5]}{Aranson I, Golomb D and Sompolinsky H 1992}{\PRL}{68}%
         {3494}
\numrefjl{[6]}{Falcioni M, Marini Bettolo Marconi U and Vulpiani A
              1991}{\PR}{44A}{2263}
\numrefjl{[7]}{Livi R, Politi A, Ruffo S and Vulpiani A 1982}
              {J. Stat. Phys.}{46}{147}

\figures
\figcaption{$x_i(n+1)$ $vs$  $x_i(n)$ for $i=48,50,52,70$ when
            $N=100$ and the defect is localized at $i=50$.
            $\ga _1=0.7$, $\ga _2=0.01$.}

\figcaption{$\ztq$  $vs$  $i$ for $j=1$ $(a)$, $j=2$ $(b)$ and
           $j=3$ $(c)$; $N=300$ and two defects are present at
           $i=70$ and $i=220$. The evolution time is $T=50 000$.}

\figcaption{$H^{(1)} - \ln N$  $vs$  $\rho$ for
            $N=200,300,500,800,1000$ }

\figcaption{Higher positive Lyapunov exponents, $\lambda_i$,
            $vs$ $i/N$ in the cases: $N=100$ (cross),
            $N=150$ (diamond) and $N=200$ (square), with a
            constant density of defects $\rho=1/50$.}

\bye